\documentclass[journal,twoside,web]{ieeecolor}
\usepackage{generic}
\usepackage{cite}
\usepackage{amsmath,amssymb,amsfonts}
\usepackage{algorithmic}
\usepackage{graphicx}
\usepackage{algorithm,algorithmic}
\usepackage{hyperref}
\hypersetup{hidelinks=true}
\usepackage{textcomp}
\usepackage{booktabs} 
\usepackage{tabularx} 
\usepackage{caption}  
\usepackage{verbatim}
\usepackage{graphicx}
\usepackage{amsmath}
\usepackage{lineno,hyperref}
\usepackage{multirow}
\usepackage{stfloats}
\usepackage{newtxmath}   
\usepackage{arydshln}
\usepackage{array} 
\usepackage{makecell}   

\def\BibTeX{{\rm B\kern-.05em{\sc i\kern-.025em b}\kern-.08em
    T\kern-.1667em\lower.7ex\hbox{E}\kern-.125emX}}
\begin{document}
\title{DRDCAE-STGNN: An End-to-End Discrimi- native Autoencoder with Spatio-Temporal Graph Learning for Motor Imagery Classification }
\author{Yi Wang, Haodong Zhang and Hongqi Li*, \IEEEmembership{Member, IEEE}
\thanks{Manuscript received Sep 3, 2025. This work was supported by the Natural Science Basic Research Program of Shaanxi Province under Grant 2024JC-YBQN-0659 (Corresponding author: Hongqi Li.) }
\thanks{Y. Wang, H. Zhang, and H. Li are with the School of Software, Northwestern Polytechnical University, Xi’an 710729, China. (e-mail: 3220251460@bit.edu.cn, zhang\_haodong@mail.nwpu.edu.cn, lihongqi@nwpu.edu.cn)}}

\maketitle

\begin{abstract}
Motor imagery (MI) based brain-computer interfaces (BCIs) hold significant potential for assistive technologies and neurorehabilitation. However, the precise and efficient decoding of MI remains challenging due to their non-stationary nature and low signal-to-noise ratio. This paper introduces a novel end-to-end deep learning framework of Discriminative Residual Dense Convolutional Autoencoder with Spatio-Temporal Graph Neural Network (DRDCAE-STGNN) to enhance the MI feature learning and classification. Specifically, the DRDCAE module leverages residual-dense connections to learn discriminative latent representations through joint reconstruction and classification, while the STGNN module captures dynamic spatial dependencies via a learnable graph adjacency matrix and models temporal dynamics using bidirectional long short-term memory (LSTM). Extensive evaluations on BCI Competition IV 2a, 2b, and PhysioNet datasets demonstrate state-of-the-art performance, with average accuracies of 95.42\%, 97.51\%, and 90.15\%, respectively. Ablation studies confirm the contribution of each component, and interpretability analysis reveals neurophysiologically meaningful connectivity patterns. Moreover, despite its complexity, the model maintains a feasible parameter count and an inference time of 0.32 ms per sample. These results indicate that our method offers a robust, accurate, and interpretable solution for MI-EEG decoding, with strong generalizability across subjects and tasks and meeting the requirements for potential real-time BCI applications.
\end{abstract}


\section{Introduction}
\label{sec:I}
\IEEEPARstart{E}{LECTROENCEPHALOGRAM} has emerged as the most widely adopted non-invasive neural signal acquisition method for brain-computer interface (BCI) systems, owing to its portability, high temporal resolution, and relatively low cost \cite{1,2}. Among various BCI paradigms, motor imagery (MI) has garnered significant research interest due to its independence from external stimuli, relying solely on the user’s internal intention to imagine limb movements. This makes MI-BCI a promising tool for neurorehabilitation, prosthetic control, and assistive technologies \cite{3}.

Generally, MI-BCI operates by decoding EEG signals generated when a user imagines movements of specific body parts, such as the left hand, right hand, or feet. These mental tasks induce characteristic neural oscillatory patterns known as event-related desynchronization (ERD) and event-related synchronization (ERS), primarily observed over the sensorimotor cortex \cite{4,5}. Despite their physiological basis, MI-EEG signals are inherently non-stationary, exhibit low signal-to-noise ratio, and show pronounced variability across subjects and sessions \cite{6}. These characteristics render conventional machine learning approaches, which often depend on hand-crafted features (e.g., power spectral density, common spatial patterns \cite{7}, Hjorth parameters \cite{8}) and shallow classifiers like support vector machine (SVM), or linear discriminant analysis (LDA), insufficient for capturing the complex spatio-temporal dynamics essential for robust decoding.

Recent advances in deep learning have shown promise in addressing these challenges. Convolutional neural networks (CNNs) excel at extracting spatially localized features \cite{9}, while recurrent architectures such as long short-term memory (LSTM) networks are effective for modeling temporal dynamics \cite{10}. Hybrid models, including CNN-Transformer architectures, have further improved decoding performance by integrating spectral, spatial, and temporal learning \cite{11}. Con-currently, graph neural networks (GNNs) have also been introduced to explicitly model the functional connectivity between EEG electrodes, treating them as nodes in a graph structure \cite{12,13}. Notwithstanding these developments, most existing methods still face limitations. Fixed or heuristic graph structures may not adequately capture subject-specific neural interactions, while purely reconstruction-based auto-encoders often fail to learn features optimal for classification. There remains a need for an integrated framework that jointly learns discriminative representations and dynamically adapts to the spatio-temporal characteristics of MI-EEG signals.

To bridge this gap, in this work, we propose a novel end-to-end architecture termed DRDCAE-STGNN, which combines a Discriminative Residual Dense Convolutional Autoencoder with a Spatio-Temporal Graph Neural Network. Overall, the DRDCAE module enhances feature learning by deep residual and dense connections, trained with a combined reconstruction and classification loss to ensure both representational fidelity and discriminative power. The STGNN module incorporates a learnable adjacency matrix to adaptively model inter-channel dependencies and utilizes graph convolution and bidirectional LSTM to capture temporal evolution. Extensive experiments conducted on BCI Competition IV 2a, 2b and PhysioNet Motor Movement datasets demonstrate that our proposed model achieves state-of-the-art performance, superior genera-lization, and meaningful interpretability of learned brain connectivity patterns. The main contributions of this work are threefold: 1) we propose a novel discriminative autoencoder enriched with residual and dense connections to learn compact and class-relevant feature representations from MI-EEG signals; 2) a dynamic graph-based STGNN architecture is ingeniously adopted to effectively model spatial and temporal dependencies; and 3) comprehensive experiments and ablation studies confirm the superiority and interpretability of our method for MI-EEG classification tasks.

The remainder of this paper is structured as follows. Section \ref{sec:II} reviews recent related work. Section \ref{sec:III} details the proposed methodology. Section \ref{sec:IV} describes the datasets and experi-mental setup. Comparative results are presented in Section \ref{sec:V}, and Section \ref{sec:VI} concludes the paper with discussions on findings and future research directions.

\section{Related Works}
\label{sec:II}
The classification of EEG signals has traditionally followed a pipeline involving signal preprocessing, feature extraction, and classification \cite{14}. Each stage has seen extensive research and methodological developments. In the context of MI-EEG classification, spatial and temporal representations are of primary importance, as they directly capture brain-region activations and task-related dynamics. By contrast, frequency-domain features, while widely applied, reflect properties common to many physiological signals and are thus regarded as auxiliary rather than central in this setting. Recent advances have shifted toward deep learning and graph-based models, which can learn representations directly from raw or minimally processed data. This section reviews related work in four key areas: spatial feature learning, temporal modeling, spatio-temporal fusion, and deep feature learning via residual or dense connections.

\subsection{Spatial Feature Learning}
Spatial modeling plays a central role in decoding EEG signals, especially for the motor imagery signals, as it aims to capture inter-channel interactions that reflect cortical connecti-vity. Graph-based approaches have gained prominence for their ability to explicitly model relationships between EEG electrodes. Kipf and Welling’s Graph Convolutional Network (GCN) laid the foundation for spectral graph learning \cite{15}, and was quickly adopted in EEG applications such as EEG-GCN \cite{12}, where predefined sensor adjacency was used to model spatial dependencies. However, such fixed graphs may not generalize across subjects or tasks.

To overcome this limitation, attention mechanisms were introduced to dynamically adjust graph connectivity. Graph Attention Network (GAT) \cite{16} learns edge weights via self-attention, allowing to dynamically adjust the graph topology. More recent approaches treat the adjacency matrix as entirely learnable, enabling task-adaptive graph construction to better capture dynamic relationships between nodes \cite{17}. Studies \cite{18} and \cite{19} further advanced this idea by introducing cross-frequency and multi-scale graph learning, demonstrating that adaptive graph structures significantly improve decoding accuracy and generalizability compared to static counterparts.

\subsection{Temporal Dependency Modeling}
EEG signals are inherently temporal, and capturing their dynamics is essential for decoding motor intentions. Early approaches combined spatial encoding with recurrent modules. EEGNet, originally proposed for compact CNN-based EEG decoding, was extended with LSTM layers to form EEGNet-LSTM, enabling sequence modeling of MI signals \cite{20}. Similarly, DeepConvNet-LSTM integrated convolutional and recurrent layers to capture temporal dependencies in P300 and epilepsy detection \cite{21}. Despite their effectiveness, these recurrent neural networks (RNNs) are often computationally intensive and prone to vanishing gradients. Temporal convolu-tional networks (TCNs) such as TS-TCN offer an alternative with larger receptive fields and more stable training \cite{22}. More recently, Transformer-based models have been applied to EEG decoding, leveraging self-attention to capture long-range temporal dependencies. Specifically, EEG-Transformer \cite{23} and multi-branch fusion Transformers \cite{24} have shown remarkable performance by attending to globally relevant time steps, marking a shift from local recurrent modeling to global temporal attention.

\subsection{Fusion of Spatial-Temporal Features}
Joint modeling of spatial and temporal features has proven highly effective across various EEG paradigms, where the spatio-temporal graph neural networks (STGNNs) have been particularly successful, combining graph convolutions for spatial modeling with sequential modules (e.g., LSTMs or TCNs) for temporal processing \cite{25,26}. For example, ST-SCGNN \cite{27}dynamically constructs graph structures for cross-subject emotion recognition and consciousness detection, demonstrating the value of adaptive spatio-temporal modeling. Similar approaches have been applied to MI-EEG, where models such as EEG-STGNN and BrainGNN \cite{28} use gated recurrent units and attention mechanisms to model evolving spatial interactions over time. In addition, Yan et al. \cite{29} proposed a temporal-spatial embedding and dynamic aggre-gation network for EEG motion intention recognition, while Zheng et al. \cite{30} introduced a deep multi-dilation model for 3D EEG. These studies consistently show that integrated spatio-temporal models outperform methods that treat spatial and temporal features independently.

\subsection{Residual and Dense Connection-based Research}
To improve robustness under the noisy and low-resource conditions, autoencoders have been widely employed for unsupervised EEG feature extraction. However, standard autoencoders often lack discriminative power. To address this, hybrid architectures are proposed that combine reconstruction loss with supervised learning loss. More specifically, deep residual connections, as popularized by ResNet, help mitigate the vanishing gradient problem and enable training of deeper networks. Discriminative convolutional autoencoders (DCAEs) \cite{31} further enhance latent feature quality through auxiliary classification loss or conditional guidance. Variational auto-encoders (VAEs) and generative adversarial networks (GANs) have also been used for EEG data augmentation and domain adaptation \cite{32}, which learn subject-invariant features by introducing probabilistic priors or adversarial training. These strategies, combined with dense connectivity, enable efficient and hierarchical feature compression, forming a strong foundation for downstream decoding tasks.

\subsection{Research Gap and Contribution}
Despite these advancements, most existing methods exhibit limitations in one or more aspects. Spatial models often rely on static graphs, temporal models may overlook spatial interactions, and autoencoders may not optimize features for classification. Moreover, few frameworks seamlessly integrate discriminative feature learning with dynamic spatio-temporal modeling, and most existing approaches often degrade under subject variability and noise, leaving room for improvement in both discriminability and generalization. 
Concluding the current advances, discriminative residual dense convolutional autoencoder (DRDCAE) excels at robust feature compression but lacks explicit modeling of complex spatio-temporal dependencies. STGNN is able to effectively capture dynamic relations while remaining vulnerable in terms of discriminative power and noise resistance. This motivates the integration of the two into a unified framework that combines discriminative feature learning with adaptive spatio-temporal modeling for MI-EEG classification. Our proposed framework follows this principle by unifying residual-dense encoding with classification-aware learning objectives.

\section{Approach of DRDCAE-STGN Network}
\label{sec:III}

The overall structure of the proposed DRDCAE-STGNN framework is illustrated in Fig. \ref{fig_1}, where the model accepts the preprocessed three-dimensional EEG data (samples × channels × features) containing both temporal-spectral and spatial features. Following standardization (i.e., Step 1 - 4) depicted in the middle part of Fig. \ref{fig_1}, the data is fed into the model. The backbone network mainly consists of two parts: discriminative residual dense convolutional autoencoder (DRDCAE) and the spatio-temporal graph neural network (STGNN), which are used for high-dimensional extraction of local features and global modeling across channels and time.
\subsection{DRDCAE Module}
The DRDCAE is designed to learn compact and discrimi-native feature representations through a combination of reconstruction and classification objectives, and consists of an encoder, decoder and classifier (see in Fig. \ref{fig_1}). Unlike conventional autoencoders, it incorporates residual learning and dense connectivity to enhance gradient flow and feature reuse across layers.

Specifically, in the DRDCAE encoder module, the local features of each channel are first extracted through the densely connected DenseBlock. This module employs a series of 1D convolution operations, connecting all historical feature maps to enhance the feature expression capability and the stability of gradient propagation. The computation in each residual block can be formulated as:

\begin{equation}
y_l = \operatorname{ReLU}\!\big( K_l \big( y_{l-1} + y_{l-2} + b_l \big) \big)
\label{eq}
\end{equation}

where $k_l$ and $b_l$ represents the convolutional kernel and bias term at the $l$-th layer, respectively. $y_{l-1}$ and $y_{l-2}$ denote the feature maps from the previous two layers. The residual structure allows the network to better capture the difference between input and output by adding past features before convolution.

Then, a 1×1 convolution is used to compress the channel dimension and map the data to a fixed-dimensional hidden feature space, resulting in the encoded representation. To maintain the original feature information and strengthen the model’s reconstruction ability, DRDCAE sets up a symmetric decoder structure. The hidden representation is restored to its original dimensions using transposed convolution, and a Sigmoid activation function is introduced to generate the reconstructed output. Specifically, the decoder reconstructs the original image $\hat{x} \in\mathbb{R}^{H\times W\times C}$ from the latent representation z using transposed convolution operations, described as:

\begin{equation}
\hat{x} = \operatorname{Sigmoid}\!\big((\sum_{k=1}^{K}K^{T}_{k}z_k+b) \big) \big)
\label{eq}
\end{equation}

where $K^{l}_{t}$ denotes the transposed convolution kernel and $z_k$ represents the $k-$th partition of the latent feature map.

A critical innovation of DRDCAE is its joint optimization via a composite loss function:
\begin{equation}\mathcal L _{DRDCAE}=\mathcal L _{rec}+\lambda\mathcal L _{cls}\label{eq}\end{equation}
\begin{equation}\mathcal L _{rec}=\frac{1}{N}\sum_{i=1}^{N}\left \| x_i-\hat{x}_i  \right \|   ^2\label{eq}\end{equation}

where $\mathcal L _{rec}$ is the mean squared reconstruction error, which evaluates the discrepancy between the original input $x_i$ and its reconstruction $\hat{x}_i$, N denotes the number of samples. $\mathcal L _{cls}$ is the cross-entropy classification loss, the hyperparameter $\lambda$ balances the two objectives, ensuring the learned features are both representative and discriminative. 

\subsection{Spatio-Temporal Graph Neural Network}
After the initial feature extraction, the encoded features from the encoder are fed into the STGNN module to further model the spatial connectivity between EEG channels and the dynamic evolution of features in the time series. The STGNN is based on a trainable adjacency matrix and consists of three key components: adaptive graph convolution, bidirectional temporal modeling, and attention-based pooling.

\begin{figure*}[htpb]
\centering
\includegraphics[width=18cm]{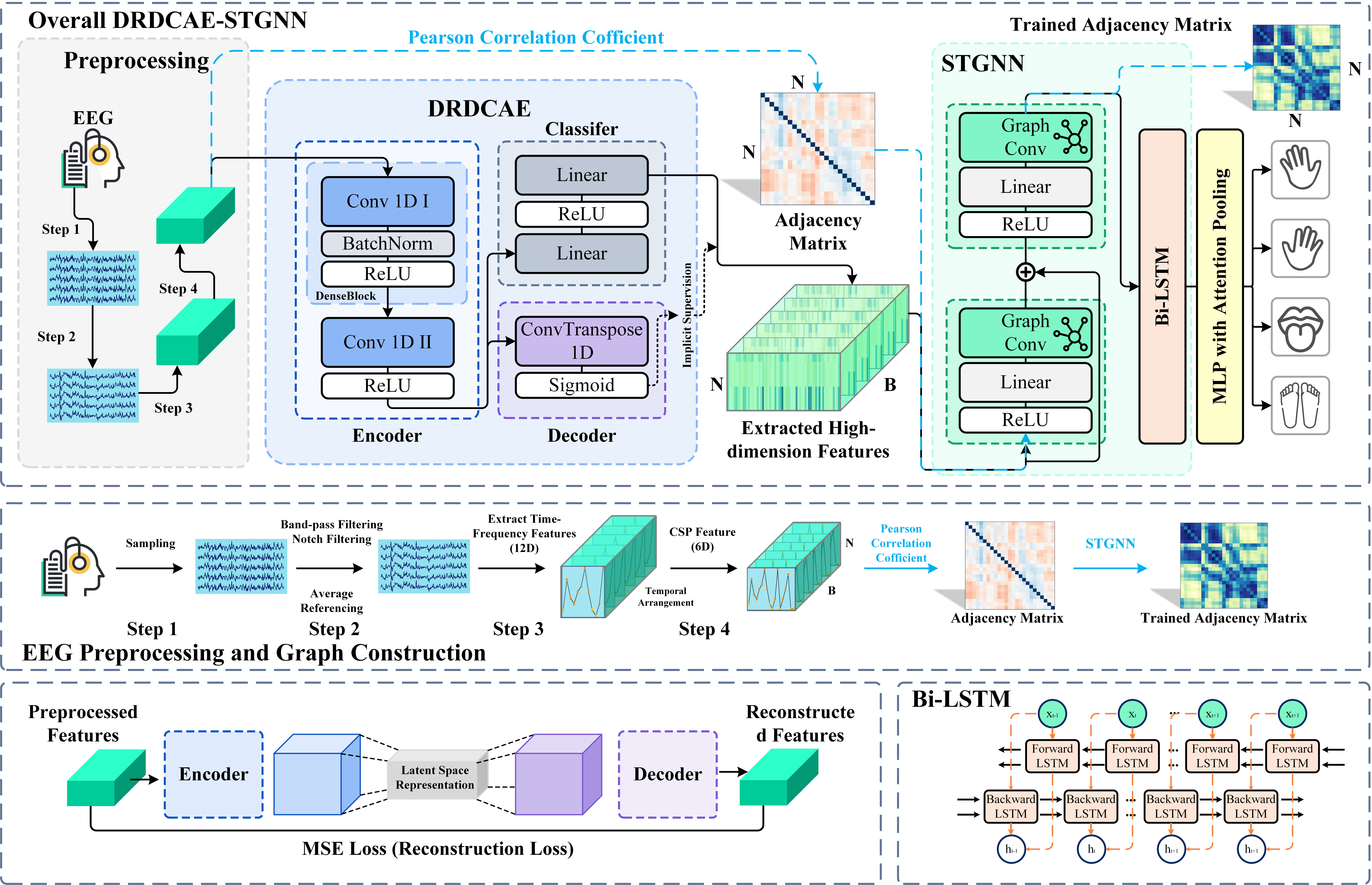}
\captionsetup{font=footnotesize}
\caption{The proposed DRDCAE-STGNN model architecture, where the overall structure is at the top, the signal processing flow diagram is in the middle, and the partial implementation details of DRDCAE and STGNN are at the bottom, respectively. }
\vspace{-1.2\baselineskip} 
\label{fig_1}
\end{figure*}

\textit{1) Data Input of Learnable Graph Construction:} To initialize the graph topology required by the model, Pearson correlation coefficients were calculated between all EEG channels based on their averaged time series, resulting in a symmetric adjacency matrix  $\mathbf{A}  \in\mathbb{R}^{C\times C}$. During training, this matrix is made learnable to adapt to subject-specific connectivity patterns. To prevent overfitting and ensure numerical stability, self-loops are masked, and the matrix is normalized row-wise via softmax: 
\begin{equation}\tilde{A} =\operatorname{softmax}(\mathbf{A} \odot \mathbf{M}  )\label{eq}\end{equation}

where \textbf{M=1-I} is a mask matrix that removes self-connections, $\odot$ denotes the element-wise (Hadamard) product. As training progresses, the model automatically adjusts the connection strengths between nodes in $\tilde{A}$ based on their semantic contribution to the classification task. This mechanism enables the transition of the graph structure from a static prior to a task-driven, learnable topology.

\textit{2) Graph Convolutional Layers:} As shown in Fig. \ref{fig_1}, we employ two graph convolutional layers to propagate spatial information across the node dimension. Each GraphConv integrates residual connections and batch normalization to mitigate the vanishing gradient problem and improve training stability. A mask is introduced in the graph convolution operation to prevent self-connections between channels. The graph convolution operation is defined as:

\begin{equation}H^{(l+1)}=\mathrm{ReLU} (\mathrm{BN}(\tilde{A}\cdot H^{(l)}+H^{(l)} )) \label{eq}\end{equation}

where $W^{l}$ and $H^{l}$ denote the learnable weight matrix and the node feature matrix at the $l$-th layer, BN(•) represents batch normalization, and ReLU is the non-linear activation function.

\textit{3) Bidirectional Temporal Modeling:} The output of the graph convolution layers is reshaped into a sequence format and processed by a bidirectional LSTM (Bi-LSTM) to capture temporal dependencies:
\begin{equation}\overrightarrow{\textbf{h}} ,\overleftarrow{\textbf{h}} = \operatorname{Bi-LSTM}(\textbf{x}_i,\overrightarrow{\textbf{h}} ,\overleftarrow{\textbf{h}})\label{eq}\end{equation}

where $  x \in\mathbb{R}^{B\times N \times H}$ is the input feature vector at time step $t$, $B$ is the batch size, $N$ is the number of nodes (EEG channels), $H$ is the hidden dimension. The forward and backward hidden states are concatenated to form a comprehensive temporal representation.

\textit{4) Attention-Based Pooling:} To extract task-relevant global representations from all temporal node embeddings, an attention-weighted pooling mechanism is introduced along the node dimension. Attention weights $\alpha_i$ are computed through a two-layer fully connected network as:

\begin{equation}\alpha_i=\frac{\operatorname{exp}(w^T_{i}\operatorname{ReLU}(w_{i}h_i))}{\sum_{j=1}^{N} \operatorname{exp}(w^T_{i}\operatorname{ReLU}(w_{j}h_j)))}  \label{eq}\end{equation}
\begin{equation}h_{\operatorname{goa}} =\sum_{i=1}^{N}\alpha_{i} \cdot h_i\label{eq}\end{equation}

Here, $w_i$ and $w_j$ are learnable weight vectors used to transform the node features $h_i$ and $h_j$, respectively. In the context of graph neural networks, $h_i$ typically represents the multi-dimensional feature vector of node $i$ (e.g., an EEG channel). The attention weight $\alpha_i$ measures the contribution of node $i$, and the resulting global representation is passed to a softmax classifier for final prediction.

\subsection{End-to-End Training}
The entire DRDCAE-STGNN model is trained in an end-to-end manner. The overall loss function combines the DRDCAE reconstruction and classification losses with the STGNN classification loss:
\begin{equation}\mathcal L _{total}=\mathcal L _{REC}+\gamma \mathcal L _{cls}\label{eq}\end{equation}

where $\gamma$ controls the contribution of the STGNN module.

In summary, the integrated architecture with adaptive graph learning strategy retains neurophysiological prior knowledge while introducing structural flexibility. It allows the model to emphasize task-relevant inter-channel dependencies and suppress irrelevant ones, thereby improving the spatial representation capacity of the graph-based feature extractor.

\section{Dataset And Experimental Setup}
\label{sec:IV}

To rigorously evaluate the performance and generalizability of the proposed DRDCAE-STGNN model, experiments were conducted on three publicly available motor imagery EEG datasets: BCI Competition IV Dataset 2a (BCI 2a), BCI Competition IV Dataset 2b (BCI 2b), and the PhysioNet EEG Motor Movement/Imagery Dataset (PhysioNet) \cite{33}. These datasets vary in complexity, number of channels, and subject pools, providing a comprehensive benchmark.

\subsection{Datasets}
\textbf{\textit{Dataset I}}: \textbf{BCI 2a}-This dataset contains EEG recordings from 9 subjects performing four-class MI tasks (left hand, right hand, feet, and tongue). Data was recorded using 22 EEG electrodes and 3 EOG channels at a sampling rate of 250 Hz. Each subject completed two sessions on separate days, each comprising 288 trials (72 per class), resulting in 576 trials per subject. In this study, all EEG channels signal was used, with the time window set for each trial between 3-6s.

\textbf{\textit{Dataset II}}: \textbf{BCI 2b}-This dataset features binary-class MI (left hand vs. right hand) from 9 subjects. Signals were recorded from 3 bipolar channels (C3, Cz, C4) at 250 Hz. Each subject participated in five sessions, with the first two without feedback and the latter three with feedback. For consistency with common practice, we utilized data from all sessions, totaling 400 trials per subject. Its lower channel count presents a distinct challenge for spatial modeling.

\textbf{\textit{Dataset III}}: \textbf{PhysioNet}-This large-scale dataset includes data from 109 subjects performing both motor execution and imagery of hand and foot movements. We focused on the two-class (left/right hand) and four-class (both hands, both feet) imagination tasks. Signals were recorded from 64 electrodes at a sampling rate of 160 Hz. 

\subsection{Data Preprocessing}
A uniform preprocessing pipeline was applied to all datasets using MNE-Python. Continuous EEG data were bandpass-filtered between 0.5-100 Hz and a 50 Hz notch filter was applied to suppress line noise. The data was then re-referenced using a common average reference (CAR). Epochs were extracted from 0 to 3 seconds relative to the cue onset. A sliding window method is then used to segment each signal for subsequent feature extraction and modeling analysis. For all datasets, trials were split into an 80/20 training-testing ratio per subject.

\subsection{Model Implementation and Training}
The proposed DRDCAE-STGNN was implemented in PyTorch. Model parameters were initialized using Kaiming initialization (for convolutional layers) and Xavier (for graph and linear layers) strategies to ensure stable training. The model was trained end-to-end using the Adam optimizer, where the learning rate and batch size are being $1e^{-3}$ and 32 for DRDCAE and $2e^{-4}$ and 16 for STGNN, respectively. Dropout (p = 0.3) was applied to mitigate overfitting. The loss balance parameters $\lambda$ and $\gamma$ were empirically set to 0.3 and 1.0, respectively. DRDCAE was trained using a weighted sum of reconstruction (MSE) and classification (cross-entropy) losses, while STGNN used cross-entropy loss. Training proceeded for a maximum of 100 epochs with early stopping based on validation loss. Detailed layer configuration and parameter counts for each component of the proposed DRDCAE-STGNN model are provided in Table\ref{tab:table1}.

\begin{table}[htpb]
    \captionsetup{font=footnotesize,justification=centering}
    \caption{\textsc{\\Layer Configuration and Parameter Counts of Proposed Model}\label{tab:table1}}
    \footnotesize
    \centering
    \begin{tabularx}{\columnwidth}{
        >{\centering\arraybackslash}p{0.1cm}
        >{\centering\arraybackslash}p{0.1cm}
        >{\centering\arraybackslash}p{0.1cm}
        >{\centering\arraybackslash}p{0.1cm}
        >{\centering\arraybackslash}p{0.8cm}
        >{\centering\arraybackslash}p{1.2cm}
        >{\centering\arraybackslash}p{2.5cm}
        >{\centering\arraybackslash}p{1.4cm}
        }
        \toprule
        \multicolumn{4}{c}{\textbf{Module}} & \textbf{Layer*} & \textbf{Input} & \textbf{Output} & \textbf{Params} \\
        \midrule
        \multicolumn{4}{c}{\multirow{5}*{DARCAE}} & DB & $(B,F,N)$ & $(B,66,N)$ & $\approx 4.8\text{K}$ \\
        \multicolumn{4}{c}{~} & Conv1D & $(B,66,N)$ & $(B,64,N)$ & $4.3\text{K}$ \\
        \multicolumn{4}{c}{~} & CT1D & $(B,64,N)$ & $(B,F,N)$ & $1.2\text{K}$ \\
        \multicolumn{4}{c}{~} & Linear & $(B,64)$ & $(B,64)$ & $4.2\text{K}$ \\
        \multicolumn{4}{c}{~} & Linear & $(B,64)$ & $(B,\text{num},\text{classes})$ & $260$ \\
        \midrule
        \multicolumn{4}{c}{\multirow{6}*{STGNN}} & GC1 & $(B,N,F)$ & $(B,N,H)$ & $0.6\text{K}$ \\
        \multicolumn{4}{c}{~} & GC2 & $(B,N,H)$ & $(B,N,H)$ & $1.0\text{K}$ \\
        \multicolumn{4}{c}{~} & Bi-L. & $(B,N,H)$ & $(B,N,H)$ & $16.6\text{K}$ \\
        \multicolumn{4}{c}{~} & Attn. & $(B,N,2H)$ & $(B,N,2H)$ & $4.2\text{K}$ \\
        \multicolumn{4}{c}{~} & Linear & $(B,2H)$ & $(B,64)$ & $4.2\text{K}$ \\
        \multicolumn{4}{c}{~} & Linear & $(B,64)$ & $(B,\text{num},\text{classes})$ & $260$ \\
        \bottomrule
    \end{tabularx}

    \caption*{\scriptsize * DB: Dense Block, CT1D: ConvTranspose 1D, Attn.: Attention, GC: Graph Convolution, Bi-L.: Bi-LSTM}
    \vspace{-0.5cm}
\end{table}

\section{Results}
\label{sec:V}
\subsection{Optimization of Sliding Window Parameters}
The selection of temporal segmentation parameters critically influences the model’s ability to capture discriminative features. We systematically evaluated the impact of window length ($\omega$) and step size (s) with their different combinations. The window length is set to 125, 250, and 500 sampling points, while the step size parameters are set to 62, 125, and 250 sampling points, resulting in a total of 9 experimental groups.

\begin{figure}[t]
\centering
\includegraphics[width=8.8cm]{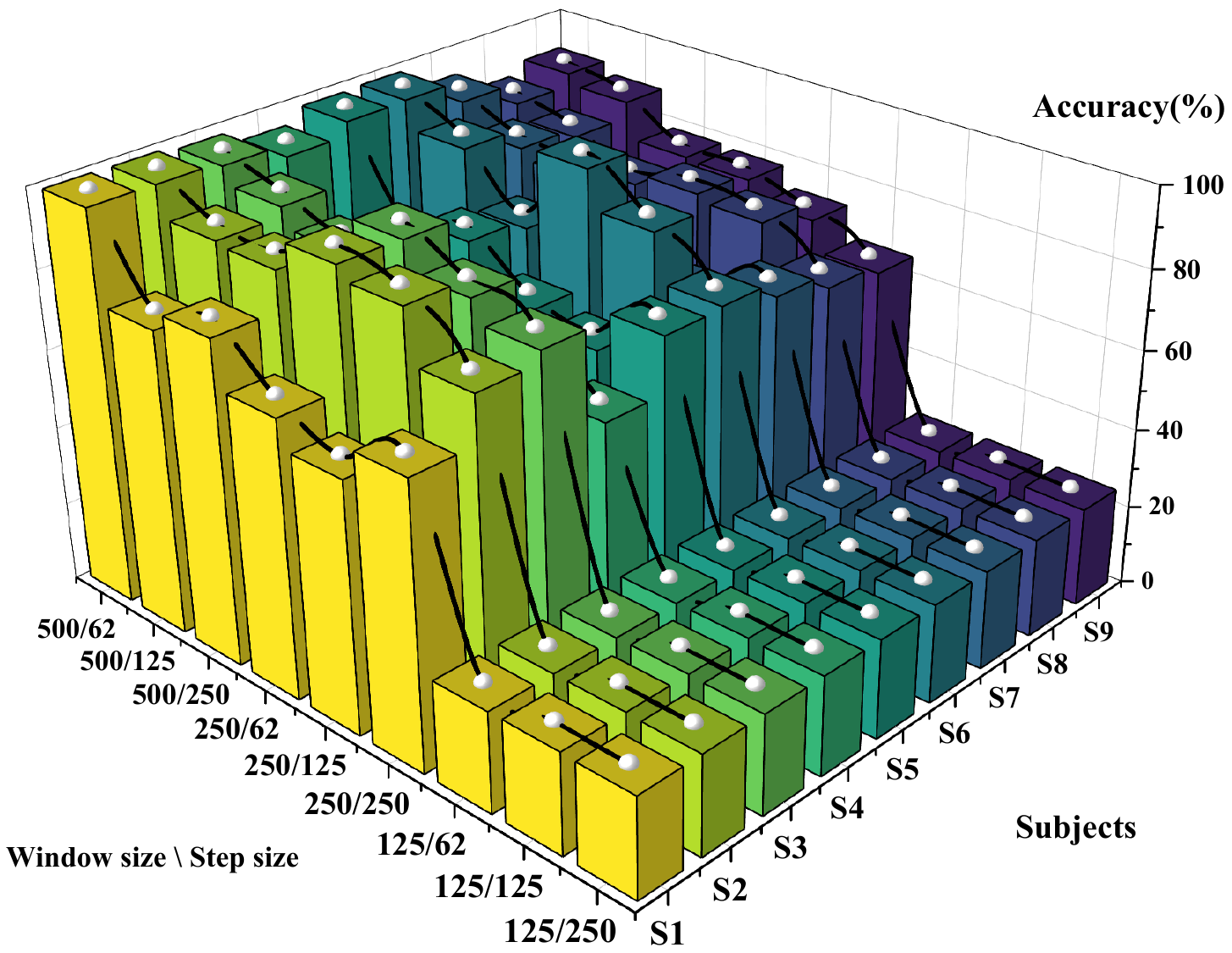}
\captionsetup{font=footnotesize}
\caption{Classification accuracy for nine subjects on the BCI-2a dataset under different window length $\omega$ and step size $s$.}
\label{fig_2}
\end{figure}

By comparing model performance, the proper selection of time resolution and feature redundancy is explored. As an example, the classification accuracy across different sliding window parameters of BCI-2a dataset is given. As illustrated in Fig.\ref{fig_2}, the configuration with a window length of 500 points (2 seconds) and a step size of 62 points (0.248 seconds) yields the highest and most stable performance across all nine subjects. Hence, this parameter set provides an optimal trade-off, where the longer window captures sufficient spectral information while the highly overlapping step ensures robustness to temporal misalignments and augments the effective training sample size. Such the configuration was consequently adopted for all subsequent experiments.

\begin{table*}[b]
    \captionsetup{font=footnotesize,justification=centering}
    \caption{\textsc{\\Classification Accuracy And Kappa Values of Different Methods on Dataset I [Avg Acc: The Average Accuracy (\%); S1~S9 Means Subject 1~9 Involved in The Dataset I; ‘-’ Indicates that the exact values were not reported by the original paper; *: P$<$0.05, **: P$<$0.01.]}\label{tab:table 2}}
    \centering
    \footnotesize
    \resizebox{\textwidth}{!}{
    \begin{tabular}{cccccccccccccc}
        \toprule
        \textbf{Year} & \textbf{Methods} & \textbf{S1} & \textbf{S2} & \textbf{S3} & \textbf{S4} & \textbf{S5} & \textbf{S6} & \textbf{S7} & \textbf{S8} & \textbf{S9} & \textbf{Avg Acc} & \textbf{Std} & \textbf{Kappa} \\
        \midrule
        2018 & EEGNet\cite{34}    & 85.71 & 78.57 & 91.15 & \textbf{95.67} & 89.20 & 85.12 & 79.23 & 81.28 & 80.67 & 85.59** & 6.02 & 0.77 \\
        2019 & SS-MEMDBF \cite{35}     & 91.49 & 60.56 & 94.16 & 76.72 & 58.52 & 68.52 & 78.57 & \textbf{97.01} & 93.85 & 79.93** & 14.99 & 0.54 \\
        2019 & MB3DC \cite{36}     & 73.97 & 60.14 & 82.93 & 72.29 & 75.84 & 68.99 & 76.00 & 76.86 & 84.67 & 75.02** & 7.29 & 0.64 \\
        2020 & JSTFD-LDA \cite{37}     & 86.40 & 55.90 & 96.30 & 73.10 & 89.50 & 58.20 & 76.10 & 93.80 & 86.60 & 79.60** & 14.78 & - \\
        2021 & DRDA \cite{38}       & 83.19 & 55.14 & 87.43 & 75.28 & 62.29 & 57.15 & 86.18 & 83.61 & 82.00 & 74.75** & 12.96 & 0.66 \\
        2022 & FBRTS \cite{39}       & 86.10 & 65.20 & 90.00 & 63.80 & 75.60 & 52.40 & 91.10 & 89.00 & 86.50 & 77.70* & 14.14 & 0.71 \\
        2023 & ATCNet \cite{40}       & 88.50 & 70.50 & 97.60 & 81.00 & 83.00 & 73.60 & 93.10 & 90.30 & 91.00 & 85.40* & 9.08  & 0.81 \\
        2024 & GMM-JCSFE \cite{41}  & 89.56 & 86.28 & 87.38 & 84.70 & 85.96 & 86.75 & 84.07 & 86.44 & 85.49 & 78.96** & 1.02 & - \\
        2024 & EEGNet-tabnet \cite{42}      & 94.60 & 79.30 & 96.70 & 86.10 & 84.20 & 78.80 & 96.20 & 90.20 & 90.40 & 92.47 & 6.83 & 0.88 \\
        2025 & TriNet \cite{43}  & 90.10 & 75.30 & 95.40 & 84.00 & 86.00 & 79.00 & 94.80 & 89.60 & 91.80 & 88.50* & 6.90 & 0.86 \\
        2025 & ACNN-STADP \cite{44}     & 90.01 & 77.00 & 97.10 & 88.70 & 88.90 & 84.70 & \textbf{98.60} & 95.40 & \textbf{94.50} & 90.54 & 6.81 & 0.87 \\
        2025 & EEG-DG \cite{45}    & 89.24 & 64.93 & 94.79 & 85.76 & 68.75 & 61.46 & 95.14 & 88.89 & 87.15 & 81.79* & 13.06  & 0.75 \\
        2025 & \textbf{DRDCAE-STGNN}       & \textbf{99.13} & \textbf{98.96} & \textbf{97.92} & 94.79 & \textbf{98.26} & \textbf{98.57} & 93.06 & 87.50 & 90.62 & \textbf{95.42} & \textbf{4.22} &\textbf{ 0.92} \\
        \bottomrule
    \end{tabular}
    }
\end{table*}

\begin{table*}[htbp]
    \captionsetup{font=footnotesize,justification=centering}
    \caption{\textsc{\\Comparison Results Of Different Methods On Dataset II [Avg Acc: The Average Accuracy(\%)  S1~S9 Means Subject 1$~$9 Involved in The Dataset II; ‘-’ Indicates that the exact values were not reported by the original paper; *: P$<$0.05, **: P$<$0.01.]}
    \label{tab:table 3}}
    \centering
    \footnotesize
    \resizebox{\textwidth}{!}{
    \begin{tabular}{cccccccccccccc}
        \toprule 
        \textbf{Year} & \textbf{Methods} & \textbf{S1} & \textbf{S2} & \textbf{S3} & \textbf{S4} & \textbf{S5} & \textbf{S6} & \textbf{S7} & \textbf{S8} & \textbf{S9} & \textbf{Avg Acc} & \textbf{Std} & \textbf{Kappa} \\
        \midrule
        2018 & EEGNet \cite{34}     & 75.94 & 57.64 & 58.43 & 98.13 & 81.25 & 88.75 & 84.06 & 93.44 & 89.69 & 80.48** & 14.40 & 0.61 \\
        2021 & FBCaps Net \cite{46}       & 79.47 & 58.34 & 59.59 & 96.40 & 84.06 & 88.09 & 82.12 & 90.47 & 89.12 & 80.85** & 13.36 & - \\
        2021 & DRDA \cite{38}       & 81.37 & 62.86 & 63.63 & 95.94 & 93.56 & 88.19 & 85.00 & 95.25 & 90.00 & 83,98* & 12.67  & 0.68 \\
        2022 & FBRTS \cite{39}       & 82.40 & 75.20 & 86.90 & 95.20 & 89.70 & 80.20 & 90.50 & 91.20 & 91.10 & 86.90** & 6.40 & - \\
        2023 & Conformer \cite{47}  & 82.50 & 65.71 & 63.75 & \textbf{98.44} & 86.56 & 90.31 & 87.81 & 94.38 & 92.19 & 84.63 & 11.49 & 0.6926 \\
        2024 & GMM-JCSFE \cite{41}    & 87.53 & 91.55 & 86.08 & 94.81& 90.58 &	90.31 &	88.76 &	87.21 &	89.90 &	89.63** &	2.64 	&- \\
        2025 & Trinet \cite{43} &93.65	&90.43 &	91.70 	&95.43 &	96.09 &	92.26 	&90.48 	&88.67 & 95.08 	&92.64** &	2.57 	&- \\
        2025 & EEG-DG \cite{17}  &  82.5&	67.50 &	72.19 &	98.44 &	96.56 &	90.94 	&89.38 &	95.00 &	91.56 &	87.12* &	10.89 &	0.74  \\
        2025 & DRDCAE-STGNN & 98.5&	97.5	&98.25 	&95.95 &	96.43 &	97.00 	&98.25 	&97.50 	&98.25 &	97.51 &	0.90 	&0.96  \\
        \bottomrule
    \end{tabular}
    }
\end{table*}

\subsection{Comparative Performance Evaluation}
The proposed DRDCAE-STGNN framework was evaluated against state-of-the-art methods on three public benchmarks. Results, detailed in TABLE \ref{tab:table 2}, TABLE \ref{tab:table 3}, and TABLE \ref{tab:table 4}, demonstrate its superior and robust performance. 
Specifically, on BCI IV-2a dataset (4-class), the proposed model achieved an average accuracy of 95.42\% ($\kappa$ = 0.92, more than half of the subjects had an accuracy rate higher than 97\%), significantly outperforming comparable methods of EEGNET \cite{34}, MB3DC \cite{36}, DRDA \cite{38}, GMM-JCSFE \cite{41} (p $<$ 0.01), and SS-MEMDBF \cite{35}, JSTFD-LDA \cite{37}, FBRTS \cite{39}, ATCnet \cite{40}, TriNET \cite{43}, EEG-DG \cite{45} (p $<$ 0.05). It also surpassed recent benchmarks such as EEGNET-tabnet (92.47\%) \cite{42} and ACNN-STADP (90.54\%) \cite{44}, while exhibiting the lower standard deviation (4.22), indicating exceptional consistency across subjects (S1-S9).

\begin{table*}[htbp]
    \captionsetup{font=footnotesize,justification=centering}
    \caption{\textsc{\\Classification Accuracy And Kappa Values of Different Methods on Dataset III [Avg Acc: The Average Accuracy (\%); AUC: Area Under Curve}\label{tab:table 4}}
    \centering
    \footnotesize
    \resizebox{\textwidth}{!}{
    \begin{tabular}{cccccc|cccccc}
        \hline
        \textbf{Year} & \textbf{Methods} & \textbf{Avg Acc} & \textbf{F1 score} & \textbf{AUC} & \textbf{Kappa} &
        \textbf{Year} & \textbf{Methods} & \textbf{Avg Acc} & \textbf{F1 score} & \textbf{AUC} & \textbf{Kappa} \\
        \hline
        
        2018 & EEGnet\cite{48} & 76.1 & 0.758 & 0.521 & 2019 &
        2019 & 4-layer LSTM\cite{52} & 54.6 & 0.526 & 0.569 & 0.090 \\
        2019& LSTM \cite{49}& 49.9	&0.494	&0.507&	-0.003&	2020&	EEGNet fusion \cite{48} & 76.2&	0.784&	0.849 &0.523\\
        2020 &	C-LSTM [50]  &74.7	 &0.736	 &0.797	 &0.495	 &
        2021	 &TS-SEFFNet [53]  &67.1	 &0.667 &	0.759	 &0.342\\
        2022 &	MBEEGNet [51]  &76.0	 &0.755	 &0.853	 &0.520	 &
        2022	 &MBEEGNet [51] & 76.5	 &0.757 &	0.846	 &0.530 \\
        2025 &	\textbf{DRDCAE-STGNN}	 &\textbf{95.9} &	\textbf{0.958}	 &\textbf{0.917} &	\textbf{0.986}	 &\textbf{2025}	 &\textbf{DRDCAE-STGNN}	 &\textbf{90.2}	 &\textbf{0.901}	 &\textbf{0.936} &	\textbf{0.868}\\

        \hline
    \end{tabular}
    }
\end{table*}

\begin{table}[htbp]
    \small 
    \captionsetup{font=footnotesize,justification=centering}
    \caption{\textsc{\\Ablation Study Results (Classification Accuracies, \%) of Four Model Variants on the BCI Competition IV-2A Dataset }\label{tab:table 5}}
    \centering
    \footnotesize
    \begin{tabularx}{\columnwidth}{>{\centering\arraybackslash}X
                                >{\centering\arraybackslash}X
                                >{\centering\arraybackslash}X
                                >{\centering\arraybackslash}X
                                >{\centering\arraybackslash}X}
        \toprule 
        \textbf{Subjects} & \textbf{Model A} & \textbf{Model B} & \textbf{Model C} & \textbf{Model D}  \\
        \midrule
        S1&	99.13&	94.35	&92.01	&85.07\\
        S2&	98.96&	96.88	&93.75	&81.94\\
        S3&	97.92	&95.14	&93.75	&78.82\\
        S4&	94.79&	91.67	&83.33	&57.64\\
        S5&	98.26&	95.49	&84.38	&72.92\\
        S6&	98.57	&95.14&	82.29	&67.01\\
        S7	&93.06	&96.18	&93.06	&71.53\\
        S8&	87.50	&91.32	&73.61	&48.96\\
        S9	&90.62	&93.06	&71.88	&44.10\\
       \textbf{ Avg Acc}	&\textbf{95.42}	&\textbf{94.36}&	\textbf{85.34}	&\textbf{67.55}\\

        \bottomrule
    \end{tabularx}
    \vspace{-0.5cm}
\end{table}

For the BCI IV-2b dataset (2-class), the model attained a mean accuracy of 97.51\% ($\kappa$ = 0.96), maintaining above 95\% for all subjects and exceeding the performance of other advanced models like Trinet \cite{43} (92.64\%), GMM-JCSFE \cite{41} (89.63\%), and EEG-DG \cite{45} (87.12\%). The remarkably low standard deviation (0.90) further confirms its robustness and adaptability to low-channel configurations. On the PhysioNet dataset, the framework demonstrated strong generalization, achieving accuracies of 95.9\% ($\kappa$ = 0.986) for the binary task and 90.2\% ($\kappa$ = 0.868) for the more challenging four-class task. The results of the F1-score and AUC also exhibit a substantial improvement over existing methods, such as MBEEGNet \cite{51}, highlighting the model’s capability to handle larger subject pools and higher channel counts. 

Overall, the proposed DRDCAE-STGNN outperforms compared baselines in both accuracy and robustness, validating the effectiveness of its residual dense encoding and spatio-temporal graph modeling.

\subsection{Ablation Study}

Ablation experiments were conducted with four model variants to quantify the contribution of each architectural component. The results on one representative dataset of BCI-2a are listed in TABLE \ref{tab:table 5}, which unequivocally validate the design choices. Specifically, the full model of DRDCAE-STGNN (Model A) achieved the highest mean accuracy (95.42\%). Replacing the STGNN with a two-layer fully connected network (Model B) resulted in a performance drop (94.36\%), underscoring the critical role of adaptive spatial modeling via graph convolutions. Removing the DRDCAE encoder and using raw time-frequency and CSP features (Model C) caused a significant decline in accuracy (85.34\%), emphasizing the importance of its discriminative feature learning and noise reduction capabilities. In addition, eliminating residual connections within the graph layers (Model D) led to the most severe performance degradation (67.55\%), confirming their necessity for stabilizing the training of deep graph networks and mitigating gradient issues.

By comparing model performance, the proper selection of time resolution and feature redundancy is explored. As an example, the classification accuracy across different sliding window parameters of BCI-2a dataset is given. As illustrated in Fig. \ref{fig_2}, the configuration with a window length of 500 points (2 seconds) and a step size of 62 points (0.248 seconds) yields the highest and most stable performance across all nine subjects. Hence, this parameter set provides an optimal trade-off, where the longer window captures sufficient spectral information while the highly overlapping step ensures robustness to temporal misalignments and augments the effective training sample size. Such the configuration was consequently adopted for all subsequent experiments.





\subsection{Analysis of Computational Efficiency}
We performed a complexity analysis with empirical evaluations to quantify the proposed model complexity, where the forward computational paths and tensor operation costs of each architecture were considered based on consistent input configurations. The running hardware configuration is AMD Ryzen 7 4800H with Radeon Graphics CPU and NVIDIA GeForce GTX 1650 Ti GPU. The results were measured on the BCI-2a dataset and were benchmarked against several classical models, as reflected in TABLE \ref{tab:table6}. As we can observe, while the integrated architecture of the DRDCAE-STGNN (with the coupled residual and dense connections, the graph convolution, bidirectional temporal modeling, and attention pooling) results in a longer training time (46.82 s), it maintains a moderate parameter count (41.92K). Most critically, the average inference time per sample is 0.3174 ms, well below the 100 ms threshold for real-time BCI operation. This demonstrates a favorable balance between high decoding accuracy and practical computational efficiency for potential deployment in the real world.

\newcolumntype{P}[1]{>{\centering\arraybackslash}p{#1}} 
\newcolumntype{Y}{>{\centering\arraybackslash}X}        
\newcommand{\hdrtwo}[2]{\shortstack{\textbf{#1}\\[-0.2ex]\textbf{#2}}}
\newcommand{\hdrempty}[1]{\shortstack{\textbf{#1}\\[-1.4ex]\phantom{Time (s)}}}

\begin{table}[htbp]
  \small
  \captionsetup{font=footnotesize,justification=centering}
  \caption{\textsc{\\Comparison of Efficiency and Complexity for Various Models}}
  \label{tab:table6}
  \centering
  \footnotesize

  \setlength{\tabcolsep}{4pt}
  \begin{tabularx}{\columnwidth}{
    @{}
    P{\dimexpr .34\columnwidth - 2\tabcolsep\relax}  
    P{\dimexpr .18\columnwidth - 2\tabcolsep\relax}  
    Y                                                
    Y                                                
    @{}
  }
    \toprule
    \hdrempty{Models} &
    \hdrempty{Parameter} &
    \hdrtwo{Training}{Time (s)} &
    \hdrtwo{Inference}{Time $(10^{-2}\,\mathrm{ms})$} \\
    \midrule
    SVM                 & 0         & 1.2   & --   \\
    CNN-LSTM            & 97.46 K   & 18.86 & 5.75 \\
    CNN-GNN             & 16.99 K   & 13.12 & 4.92 \\
    DCGAN-D \cite{54}   & 432.97 K  & 9.13  & 3.15 \\
    \mbox{\textbf{DRDCAE-STGNN}} & 41.92 K   & 46.82 & 31.74 \\
    \bottomrule
  \end{tabularx}
\end{table}

\subsection{Interpretability of Learned Graph Structures}

\begin{figure}[htbp]
\centering
\includegraphics[width=8.8cm]{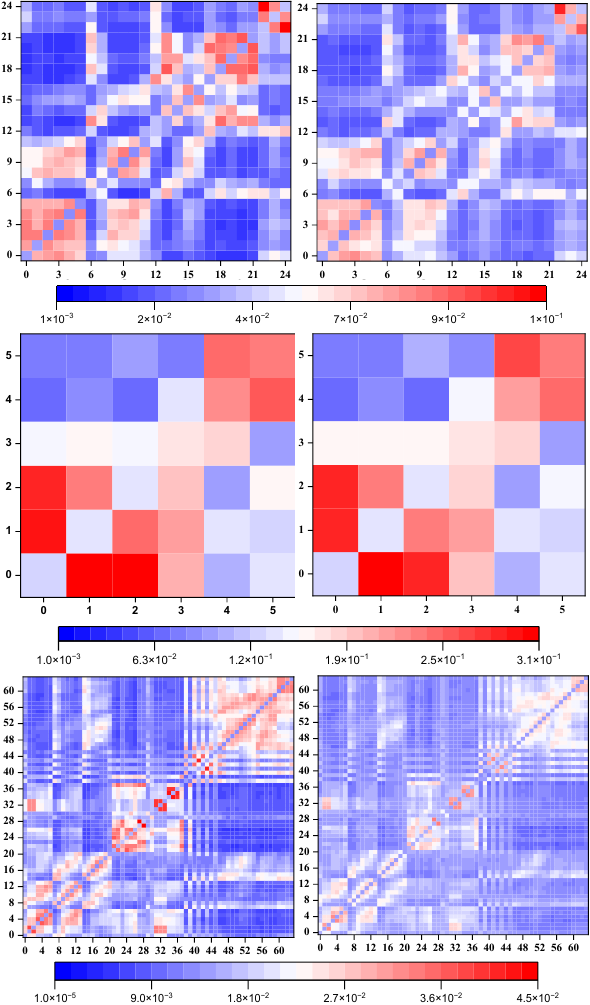}
\captionsetup{font=footnotesize}
\caption{Channel connectivity adjacency matrices before (left) and after (right) training across three datasets of BCI2a (top), BCI2b (middle), and PhysioNet (bottom). The coordinate numbers correspond to the EEG electrode under the International 10-20 system, and the color maps represent the connection strength between EEG channels.}
\label{fig_3}
\end{figure}

The learnable adjacency matrix, initialized from raw EEG features, is dynamically updated during training to capture task-relevant inter-channel dependencies and thus providing a window into the model’s spatial reasoning. Fig. \ref{fig_3} visualizes the graph connectivity before and after training across all three datasets. As we can see, the initial adjacency matrices after Softmax normalization exhibit sparse and low-weight connec-tions (more deep blue parts that indicating weaker relation), with only moderate symmetric correlations between adjacent channels, reflecting the high noise and non-stationarity of EEG signals. In contrast, post-training matrices reveal that the model consistently enhanced physiologically meaningful connections. Especially, considering the exact EEG electrode location, it formed clusters between frontal-parietal and sensorimotor regions, while suppressing spurious or noisy links. This indicates an adaptive learning process that prioritizes motor imagery task-relevant neural pathways.

\begin{figure}[!t]
\centering
\includegraphics[width=8.8cm]{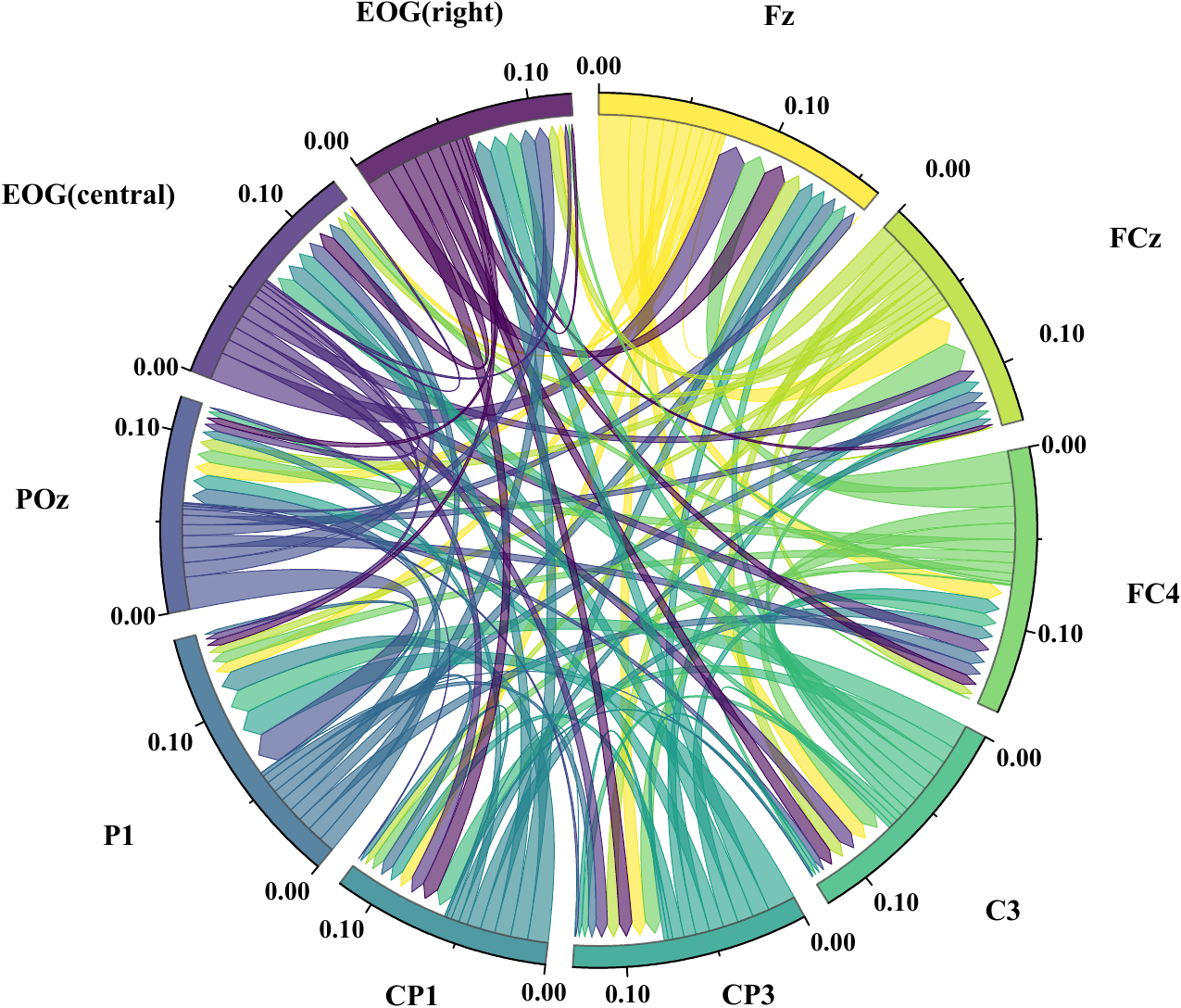}
\captionsetup{font=footnotesize}
\caption{Chord diagram of the top 10 EEG channel pairs with the largest changes in connection strength.}
\label{fig_4}
\end{figure}

Further analysis via the chord diagram (Fig. \ref{fig_4}) identifies the top 10 connections that underwent the largest strengthening during training on dataset I of BCI-2a. The prominence of links involving central (e.g., C3), parietal (e.g., CP1, P1), and frontal (e.g., Fz, FC4) electrodes aligns perfectly with the neuroanatomical correlates of motor imagery, providing strong evidence for the neurophysiological validity and interpretability of the model’s learned representations.

\section{Discussion and Conclusion}
\label{sec:VI}
\subsection{Discussion of the Results and Research Contributions}

To further improve the decoding effect while maintain the interpretability, a novel DRDCAE-STGNN framework has been proposed in this study. The consistently superior performance of the framework across three distinct datasets underscores its efficacy as a powerful tool for MI-EEG decoding. The results affirm that the integration of discriminative feature learning (via DRDCAE) with adaptive spatio-temporal modeling (via STGNN) addresses core challenges in the field: the non-stationarity of EEG signals and the high variability between subjects.

The key innovation of this work is the end-to-end learning of a neuro-physiologically plausible graph structure. Unlike methods relying on static, anatomically predefined graphs, our learnable adjacency matrix allowed the model to discover and emphasize task-specific functional connectivity. The post-hoc analysis (Fig. \ref{fig_3}, Fig. \ref{fig_4}) revealed that the model autonomously strengthened connections within well-established sensorimotor networks (e.g., between frontal, central, and parietal regions) while suppressing irrelevant noise. This capability transforms the model from a black-box classifier into a more transparent tool that can offer insights into the underlying neural mechanisms of MI. Importantly, without explicit anatomical priors, the model automatically enhances connections between motor-related regions (e.g., Fz-FCz, FC4-FCz, Fz-FC4), consistent with classical motor imagery studies \cite{55}. Indeed, STGNN employs a soft-attention masked graph learning mechanism, enabling spatial sparsity and dynamic inter-channel dependencies, thus offering greater adaptability compared with static graphs.

The implications of this work are threefold: 1) It presents a highly accurate and generalizable model for MI-BCIs, with direct potential applications in neurorehabilitation and assistive control; 2) It advances the interpretability of deep learning models in neuroscience by demonstrating how learnable graph structures can reveal biologically meaningful connectivity; 3) It provides a versatile framework that can be adapted for other EEG-based tasks that rely on spatio-temporal feature learning.

\subsection{Limitations and Future Work}
In conclusion, we have proposed and validated the DRDCAE-STGNN, a novel deep learning framework that sets a new state-of-the-art for MI-EEG classification. Its success stems from a synergistic design: the DRDCAE module learns robust and discriminative latent representations by jointly optimizing for reconstruction and classification, while the STGNN module dynamically captures the evolving spatial relationships between EEG channels and their temporal context. Extensive experimental results on public benchmarks confirm that our method outperforms existing approaches in both accuracy and robustness. The adaptive graph learning component further provides neurophysiologically interpretable insights into functional brain connectivity patterns during motor imagery.

However, several limitations of our current work must be acknowledged. First, while the model demonstrates strong overall performance, its accuracy varies across subjects (e.g., S8 and S9 in BCI-2a). This variability suggests that the current architecture, though adaptive, may not fully compensate for the pronounced physiological and cognitive differences between individuals. Second, the computational cost during training, though justified by the performance gains, is non-trivial. This could pose a barrier for deployment in scenarios requiring rapid model personalization. Third, while the graph provides high-level interpretability, the exact contribution of specific time-points or frequency bands to the final decision remains less transparent, indicating a need for more granular explainability techniques.

Looking forward, we plan to address the current limitations through several avenues. First, to enhance the cross-subject generalization ability, we will attempt to strengthen model adaptability through subject-invariant representation learning that combines adversarial domain alignment with covariance correction on the symmetric positive definite (SPD) manifold \cite{56}. Furthermore, we will introduce parameter-efficient test-time adaptation modules to minimize calibration overhead for new users, thereby enhancing the practical utility of our framework in real-world BCI applications. Second, for the multi-granular interpretability, we will develop enhanced explanation methods that integrate gradient-based relevance analysis with perturbation testing to generate channel- and time-frequency-level attributions. These attributions will be quantitatively validated against well-established $\mu/\beta$ motor rhythm patterns \cite{57}. Finally, we will extend the framework to asynchronous ERP detection and emotion recognition within multi-task EEG paradigms, and evaluate it with rigorous leave-one-subject-out and cross-dataset validation protocols to ensure robustness and scalability. These efforts will ultimately facilitate more reliable and trustworthy BCI systems for clinical and assistive applications.

\end{document}